\documentclass[sn-mathphys]{sn-jnl}

\usepackage{amssymb}
\usepackage{graphicx}
\usepackage{epstopdf}
\usepackage{amsmath}
\usepackage{epsfig}
\usepackage{color}
\usepackage{ulem}
\usepackage{stmaryrd}

\jyear{2021}

\theoremstyle{thmstyleone}

\theoremstyle{thmstyletwo}

\theoremstyle{thmstylethree}

\begin{document}

\title[Regular and Chaotic Motion of Two Bodies Swinging on a Rod]{Regular and Chaotic Motion of Two Bodies Swinging on a Rod}

\author*[1]{\fnm{Lazare} \sur{Osmanov}}\email{lazare.osmanov1521r@gmail.com}

\author[2]{\fnm{Ramaz} \sur{Khomeriki}}\email{khomeriki@hotmail.com}

\affil*[1]{\orgdiv{School of Physics}, \orgname{Free University of Tbilisi}, \orgaddress{\street{David Aghmashenebeli Alley}, \city{Tbilisi}, \postcode{0159}, \country{Georgia}}}

\affil[2]{\orgdiv{Department of Physics}, \orgname{Ivane Javakhishvili Tbilisi State University}, \orgaddress{\street{3 Chavchavadze}, \city{Tbilisi}, \postcode{0128},  \country{Georgia}}}

\abstract{We investigate regular and chaotic dynamics of Two Bodies Swinging on a Rod, which differs from all the other mechanical analogies: depending on initial conditions, its oscillation could end very quickly and the reason is not a drag force or energy loss. We use various tools to analyze motion, such as Poincar\'e section for quasi-periodic and chaotic cases. We calculate Lyapunov characteristic exponent by different methods including Finite Time Lyapunov Exponent analysis. Our calculations show that the maximal Lyapunov exponent is always positive except in the marginal cases when one observes quasi-periodic oscillations. }

\keywords{Chaos, Coupled oscillators, Lyapunov's exponent, Poincare's section}

\maketitle

\section{Introduction}\label{sec1}

As it follows from Poincar\'e-Bendixson theorem \cite{P,B}, three first-order autonomous differential equations are enough to observe chaos, that's why chaotic motion is very common not only in complex systems covering all branches of physics \cite{c10,c12,c21,c1,c2}, but it is also observable in the systems with few degrees of freedom, e.g. in simple mechanical constructions \cite{c3,c4,c5,c6,c7,c13,c14,c15}, where the studies on double pendulum \cite{c8,ref1} is a most prominent example. The double pendulum has been deeply investigated using numerical, analytical, and experimental methods, and the computer simulations results coincide well with experimental measurements \cite{yorke}.

\begin{figure}
{\includegraphics[width=8cm]{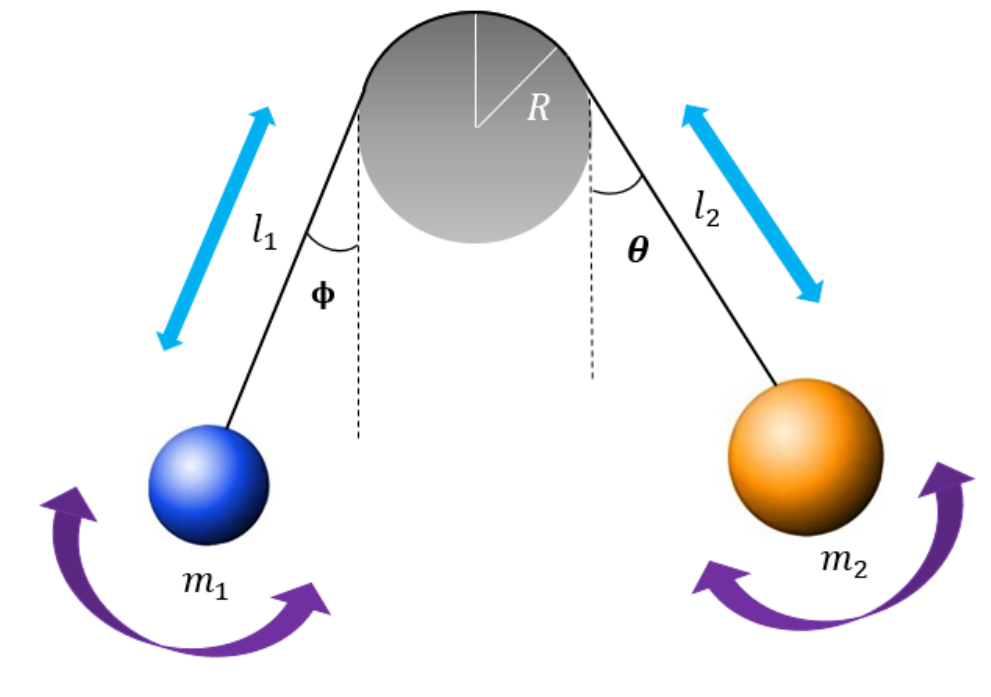}
\includegraphics[width=8.5cm]{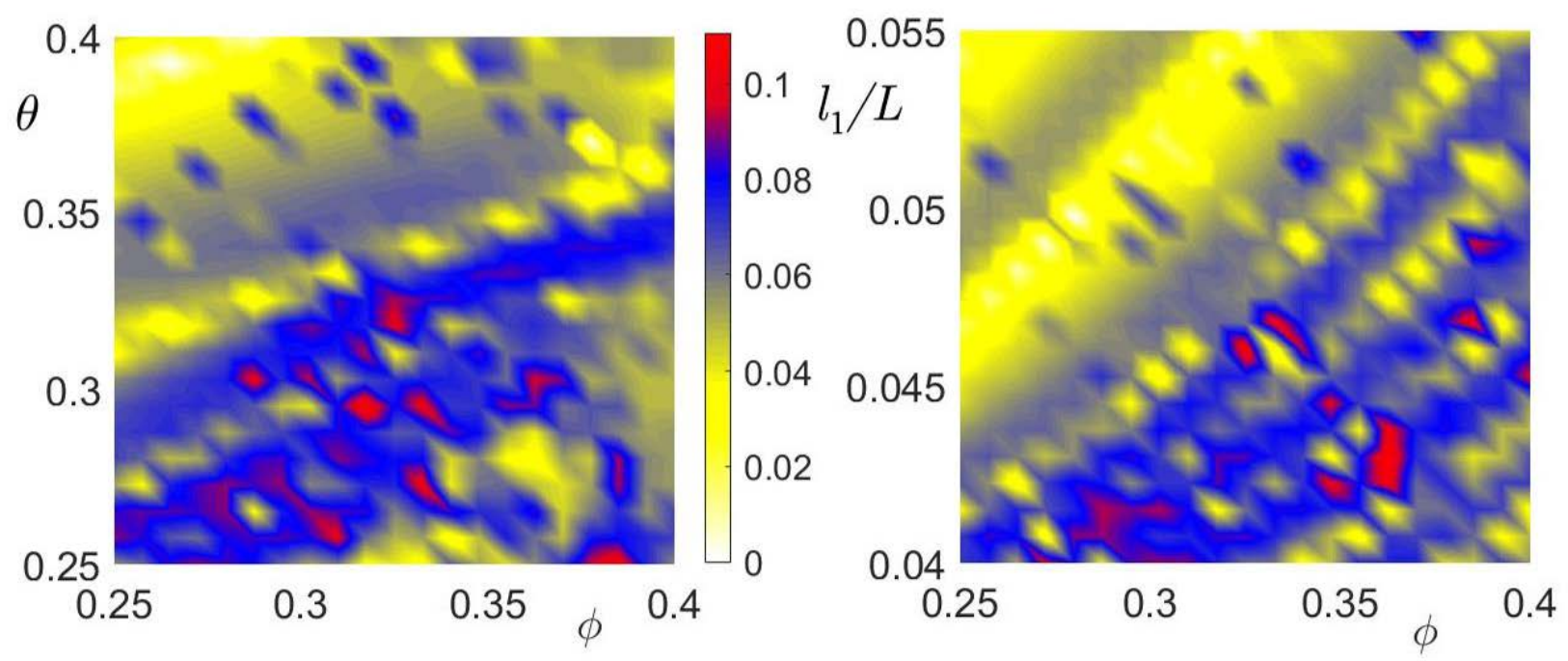}}
\caption{Upper graph: schematics for coupled fallen pendula system. $m_1$, $m_2$ are masses of the loads, and $l_1$, $l_2$ are distances between the loads and touching points of the rope with the rod. The radius of the rod is $R$. The rope can slide without friction on the rod. The arrows indicate oscillation directions of the loads,  angle $\theta$ grows in the positive, counterclockwise direction, while $phi$ in the clockwise direction. Lower graphs display density plots obtained from calculations of the largest value of Finite-time Lyapunov exponents for different initial values of $\phi$, $\theta$, and $l_1/L$. In the left graph we fix $l_1(0)/L=0.04$, while in the right graph $\theta(0)=\pi/10$ is chosen; In both cases we have fixed mass ratio as $m_1/m_2=1.1$ and Ratio of radius to full length as $R/L=1/25$ and take all initial velocities as $\dot\theta(0)=\dot\phi(0)=\dot l_1(0)=0$.} \label{fig1}
\end{figure}

Here we present another double pendulum (see the main plot in Fig. 1) having perspectives for experimental realization. Usually, chaotic motion is examined keeping in mind the infinite time limit for observation, however in our case of fallen coupled pendula the lifetime of the oscillatory motion is large in very few cases, mostly it ends at a finite time scale, so it's an example of transient chaos \cite{tc}. As far as the system is conservative, it seems that using energy as the main control parameter might be helpful, but as a result of simulations show, for the same energy level but different initial conditions we might observe two extremely different ending times.  Therefore alongside the Poincar\'e map treatment and measuring the largest Lyapunov exponent, it is desirable to examine Finite Time Lyapunov Exponent (FTLE) calculation method \cite{haller} which has been successfully applied in the study of various oscillatory systems \cite{ftle1,ftle2,ftle3} and even for determination of various regimes of incompressible and compressible flows \cite{gonzalez,3,4,5,6}. We will demonstrate that the introduction of the FTLE method in our case appears to be a very useful tool since it allows us to treat all regimes for those the lifetime is not too short. We note that the use of traditional approaches such as Poincar\'e sections or calculation of the Largest Lyapunov exponent requires large time scales for simulations, which is possible only in a few cases for our system, while for reliable measurement of FTLE the moderate time scale $\sim 100$ in dimensionless units appears to be enough.

\begin{figure}
\includegraphics[width=8.7cm]{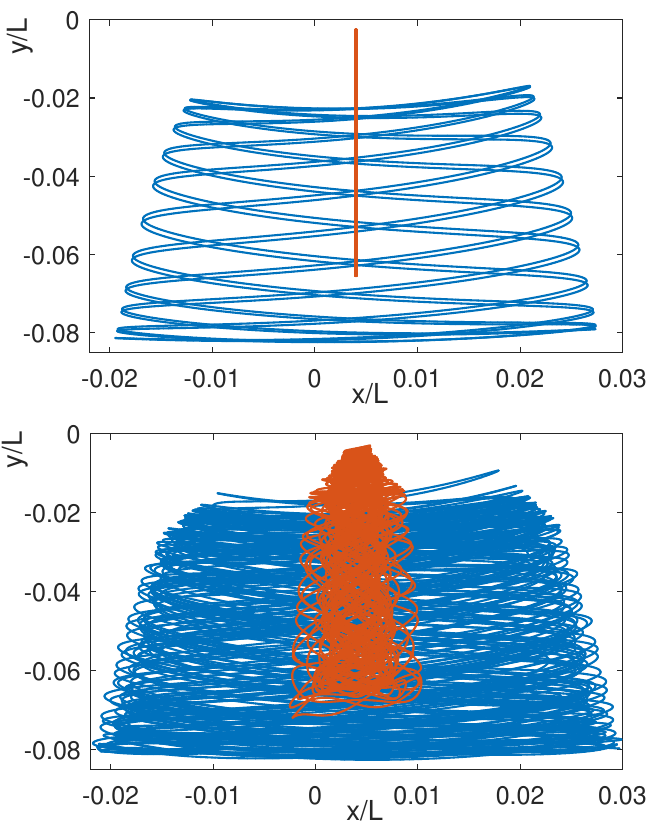}
\caption{Quasi-periodic and chaotic dynamics of the fallen pendula system. These graphs represent trajectories of both loads in different colors ($x$ and $y$ are Cartesian coordinates of the loads). In the upper graph $\theta(0)$ is taken equal to zero and in the lower graph $\theta(0)=\pi/10$. In both cases  $\phi(0)=\pi/10$, $\ell=0.05$ and all the initial velocities $\dot\phi(0)$, $\dot\theta(0)$ and $\dot\ell(0)$ have a zero value. dimensionless parameters are as it follows:$m_1/m_2=1.1$, $R/L=1/25$.}  \label{fig2}
\end{figure}

In the present paper, we consider coupled pendula system presented in Fig. 1, where instead of a pulley one has a rod with a screw-like surface to shift the oscillation planes of the two loads from each other, which, in principle, ensures the loads not to collide with each other during the oscillation process. The ropes can be wrapped completely around the rod but in our choice of initial conditions we never observed wrapping. In this nontrivial oscillatory model, unlike other mechanical systems, the motion does not last forever even in frictionless cases. This happens because oscillating loads could fall off if one of the loads reaches the rod or if the rope jumps from the rod. We also stop numerical calculations when the tension of the rope becomes zero because then our equations don't make any sense. However, in our simulations, we have never observed the latter scenario. Because of this fact, the construction of any kind of bifurcation diagram becomes an even more complicated task (for us even impossible) than in conventional conservative systems, since two infinitesimally close points in phase space can produce either quickly ending or long-lasting oscillating regimes. Thus the ultimate goal of the study is to characterize quantitatively both quickly ending and long-lasting regimes by characteristic Lyapunov exponents calculated via the FTLE method.    

Thus, below in the paper, we, first of all, investigate long-lasting oscillatory regimes of coupled pendula system by traditional methods of oscillatory systems, namely Poincar\'e maps representation is exploited and calculation of largest Lyapunov exponent is undertaken via analyzing initially infinitesimally close chaotic long-lasting trajectories. The results for such cases are compared with the FTLE method and then the latter is used for a wide range of initial conditions (irrespective of the lifetime of respective oscillatory regimes) displayed in the lower graphs of Fig. 1. Namely the density plots are presented for largest FTLE varying initial values of relative distance $l_1/L$ between first load and the rod ($L$ is a total length of the rope) and deviation angles $\theta$ and $\phi$ from equilibrium positions of the first and second loads, respectively. In both plots, we keep all initial velocities equal to zero.

\section{Equations of motion}

Equations of motion are derived using Newton's second law and we can immediately come up with the following system with the notation used in Fig. 1:
\begin{eqnarray}
T=m_2(l_2\dot{\theta}^2-\ddot{l_2}+R\ddot{\phi})+m_2g\cos{\theta} \nonumber \\
g\sin{\phi}=\dot{\phi}^2R-(\ddot{\phi}l_1+2\dot{\phi}\dot{l_1})
\nonumber \\
 g\sin{\theta}=\dot{\theta}^2R-(\ddot{\theta}l_1+2\dot{\theta}\dot{l_1}) \nonumber \\
 T-m_1g\cos{\phi}=m_1(l_1\dot{\phi}^2-\ddot{l_1}+R\ddot{\phi}) \label{a1}
\end{eqnarray}
where $T$ is the tension of the rope, $L$ is its total length, $l_1$ and $l_2$ are the distances between first and second loads and touching points of the rope with the rod, it is easy to see that $l_1$ and $l_2$ are connected via simple relation: 
\begin{equation}
\label{a5}
 l_2=L-R(\pi-\phi-\theta)-l_1. 
 \end{equation}

Let us note that it has no physical sense to continue the simulations on Eqs. \eqref{a1} if $l_1$, $l_2$ or $l_3=R(\pi-\phi-\theta)$ reach zero value ($l_3$ is a part of the rope touching the rod). In such a case we stop calculations and associate this moment with the ending time (lifetime) of the process.

Proceeding with dimensionless time $t\rightarrow t\sqrt{g/L}$ and introducing new length definitions $\ell=l_1/L$, $l\equiv l_2/L$, $r=R/L$ we get three equations for three independent variables $\phi$, $\theta$ and $\ell$:  

\begin{eqnarray}
(l\dot{\theta}^2-\ddot{l}+r\ddot{\theta})+\cos{\theta}=\frac{m_1}{m_2}\left[(\ell\dot{\phi}^2-\ddot{\ell}+r\ddot{\phi})+\cos{\phi}\right] \nonumber \\
\sin{\phi}=\dot{\phi}^2r-(\ddot{\phi}\ell+2\dot{\phi}\dot{\ell}), \quad
 \sin{\theta}=\dot{\theta}^2r-(\ddot{\theta}l+2\dot{\theta}\dot{\ell}) ~~~ \label{a11}
\end{eqnarray}
together with relation $l=1-r(\pi-\phi-\theta)-\ell$ coming from \eqref{a5}.

In simulations on Eqs. \eqref{a11} we observe four different forms of motion: the first one is the equilibrium case, which eventually happens only if $m_1/m_2=1$ and $\phi(0)=\theta(0)=0$, the second one is quickly ending cases where we can not say whether the process is chaotic or not, the third scenario is the quasi-periodic motion that never ends and might happen, for instance, in case of $m_2>m_1$ and in the small initial angle limits $\phi\lesssim 0$ and $\theta\lesssim 0.2$ (all the initial velocities taken equal to zero), and finally we observe chaotic regimes which last enough time to characterize the motion by Poincar\'e sections and Lyapunov exponent.

\begin{figure}[t]
\includegraphics[width=8.5cm]{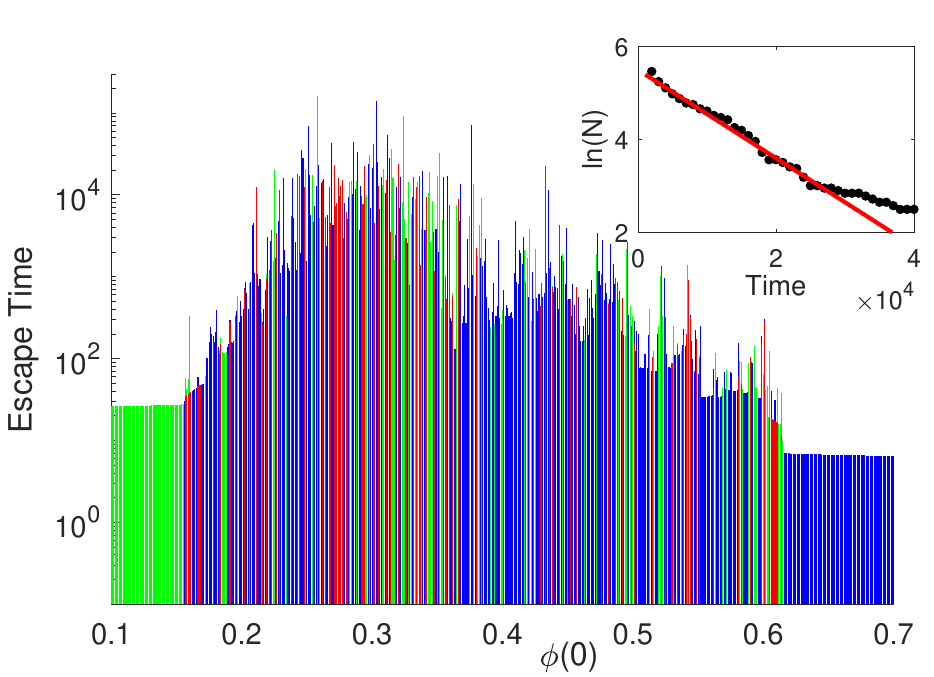}
\caption{Main plot: System lifetime (escape time) dependence on initial value $\phi(0)$. Other initial parameter values are $\ell(0)=0.05$ and $\theta(0)=\pi/10$, dimensionless parameters are: $m_1/m_2=1.1$, $R/L=1/25$. All initial values of time derivatives $\dot\phi(0)$, $\dot\theta(0)$ and $\dot\ell(0)$ are zero. By green, blue and red color are defined the escape times when either $l_1$, $l_2$ or $l_3$ becomes zero (Green, blue and red colors represent reaching of the conditions $l_1=0$, $l_2=0$ and $l_3=0$, respectively). Inset shows the estimate (slope of dashed line) of escape rate, where the time step size is taken equal to $1000$. In vertical axis we count how many trajectories survive after certain time steps and then we calculate the slope of the obtained curve. In these conditions we get escape rate value $\kappa=1/10.5$.} \label{fig3}
\end{figure}

\begin{figure}
\centering\includegraphics[width=8cm]{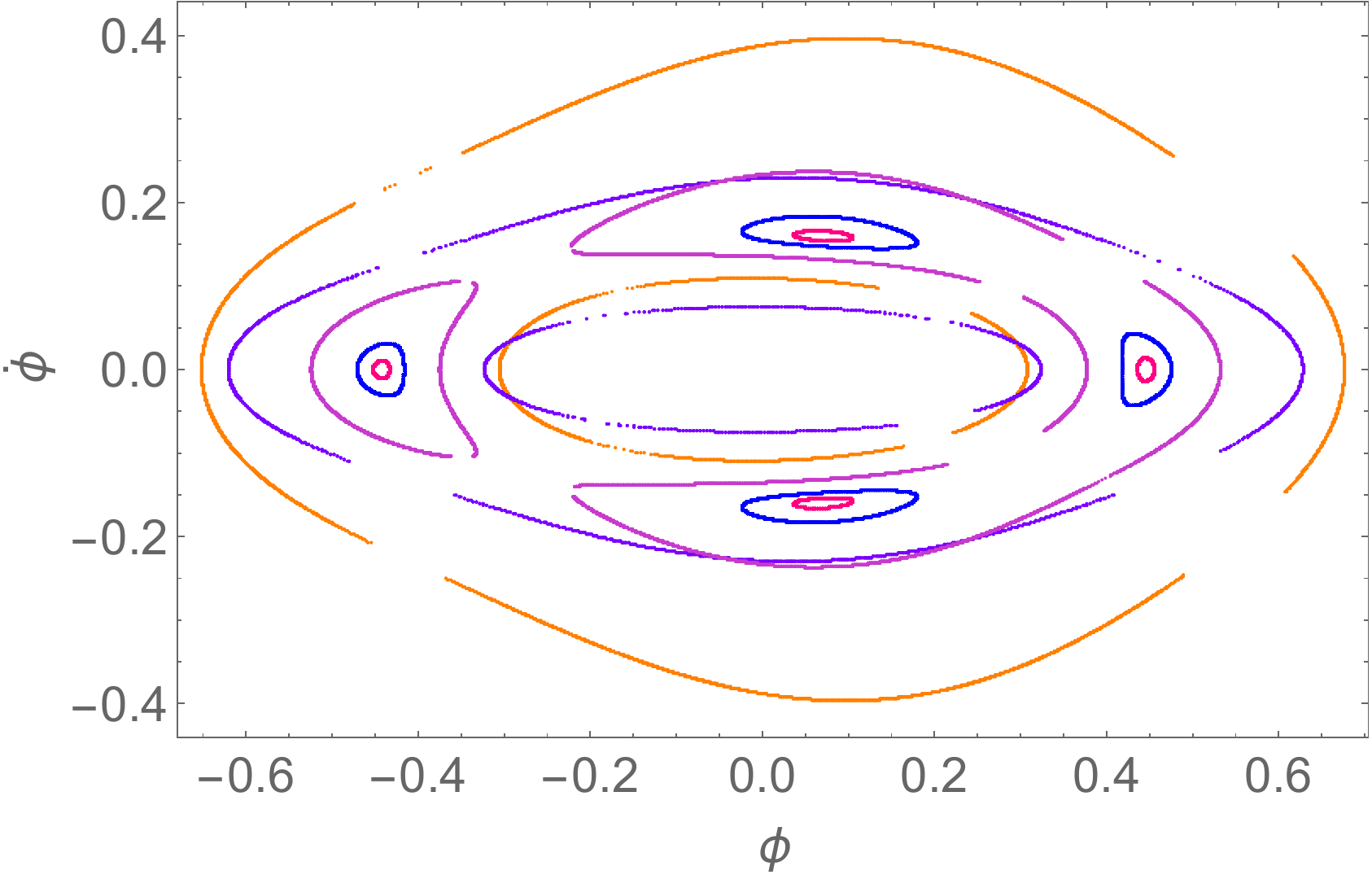}
\centering\includegraphics[width=8.15cm]{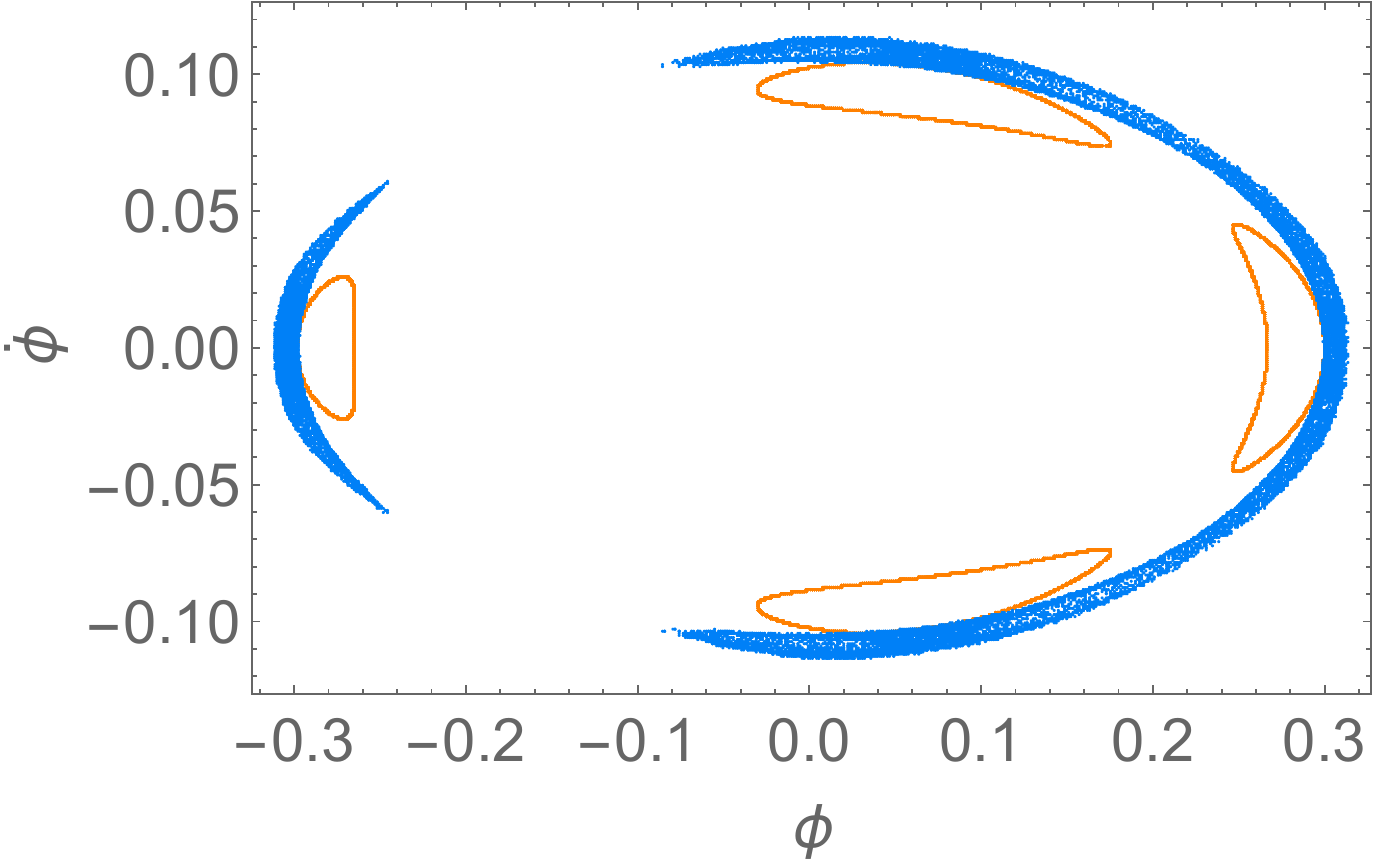}
\centering\includegraphics[width=8cm]{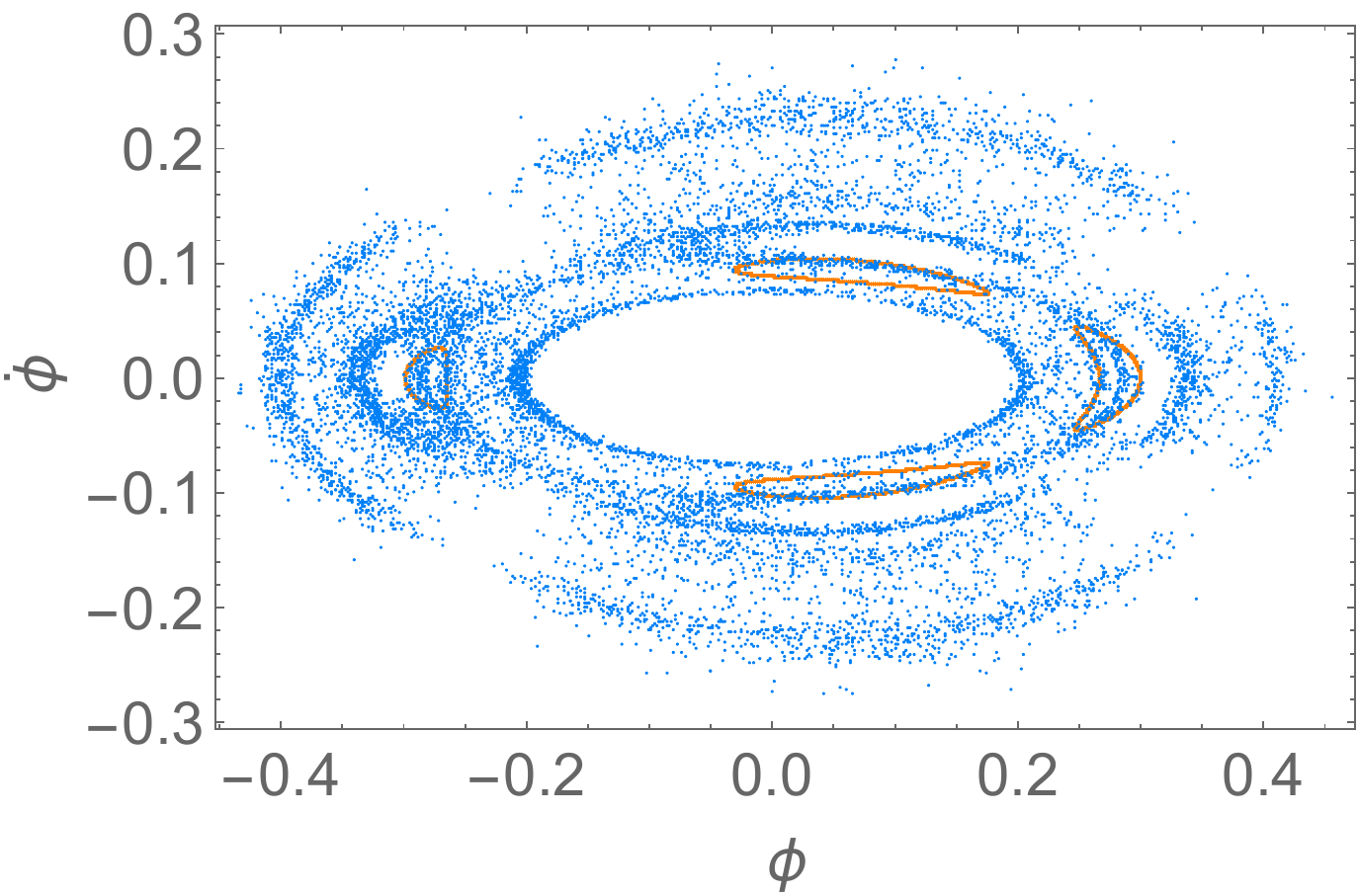}
\caption{Poincar\'e sections for Quasi-periodic and chaotic motions. Dimensionless parameters and initial values of variables for Quasi-periodic behaviour in upper graph are as follows: $m_1/m_2=1.1$, $R/L=1/25$, $\theta(0)=0$ and $\phi(0)={\pi/5,\pi/7,\pi/8,\pi/9,\pi/10}$, while in all graphs the initial values of time derivatives $\dot\phi(0)$, $\dot\theta(0)$ and $\dot\ell(0)$ we keep equal to zero. For middle graph we have slightly changed $\theta(0)$ taking it equal to $\pi/30$ and  still monitor quasi-periodic motion (orange curves we borrow from upper graph which correspond to the case $\phi(0)=0.3$ and $\theta(0)=0$). In the Lower graph with $\theta(0)=\pi/10$ the motion is strongly chaotic (note, the orange curves are still kept). Reduced initial length of the rope in all cases is $\ell(0)=0.05$}.\label{fig4}
\end{figure}

\begin{figure}
\resizebox{\hsize}{!}{\includegraphics[width=8cm]{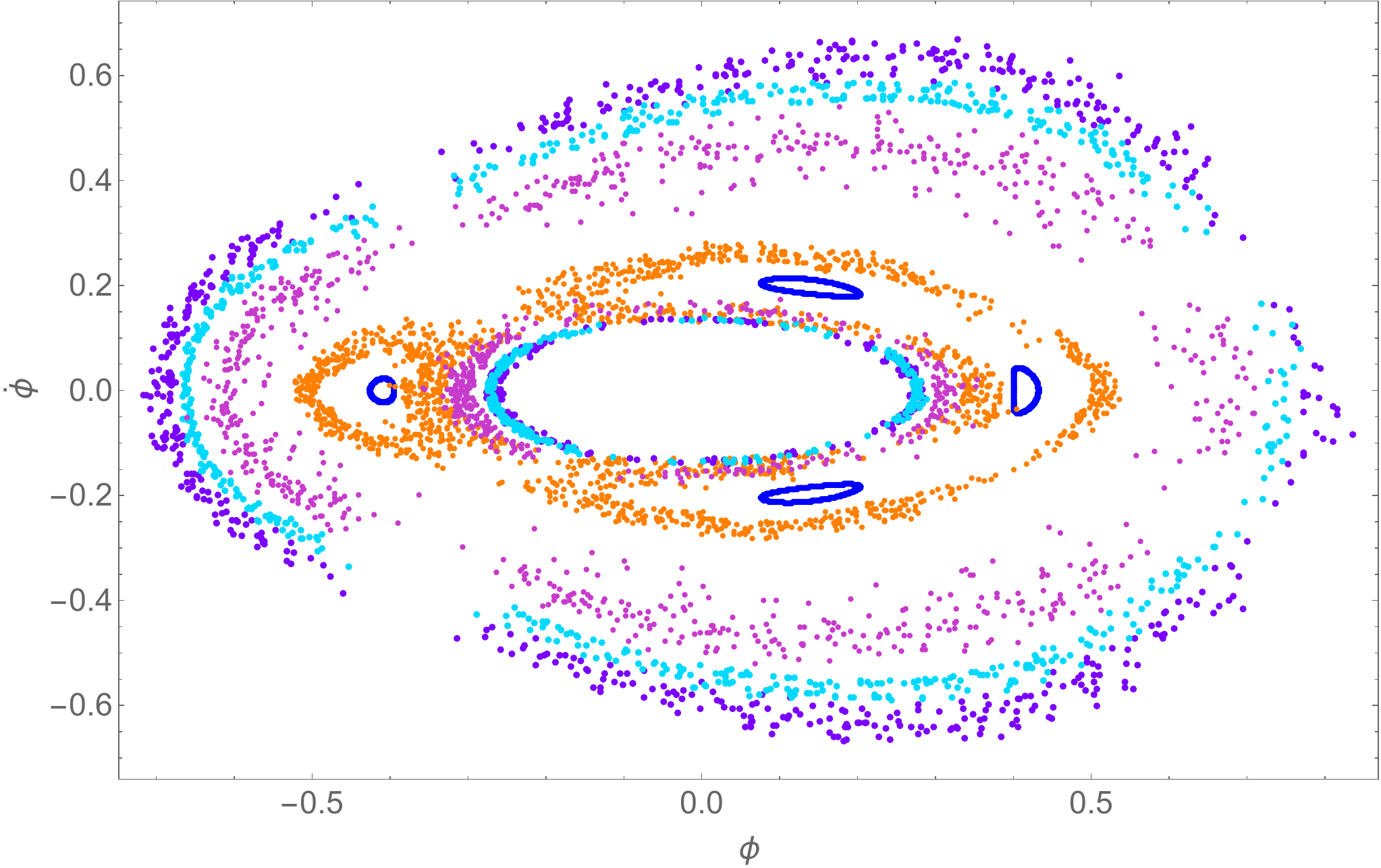}}
\caption{Here we fixed energy $E=0.304$ for the following parameters: $\theta_0=0$, $\phi_0=0.4$, $l=0.05$ and all initial time derivatives are zeros. Then, keeping the same energy, we are changing initial parameters in case of nonzero $\theta$ value.} \label{fig5}
\end{figure}

\begin{figure}
\resizebox{\hsize}{!}{\includegraphics[width=8cm]{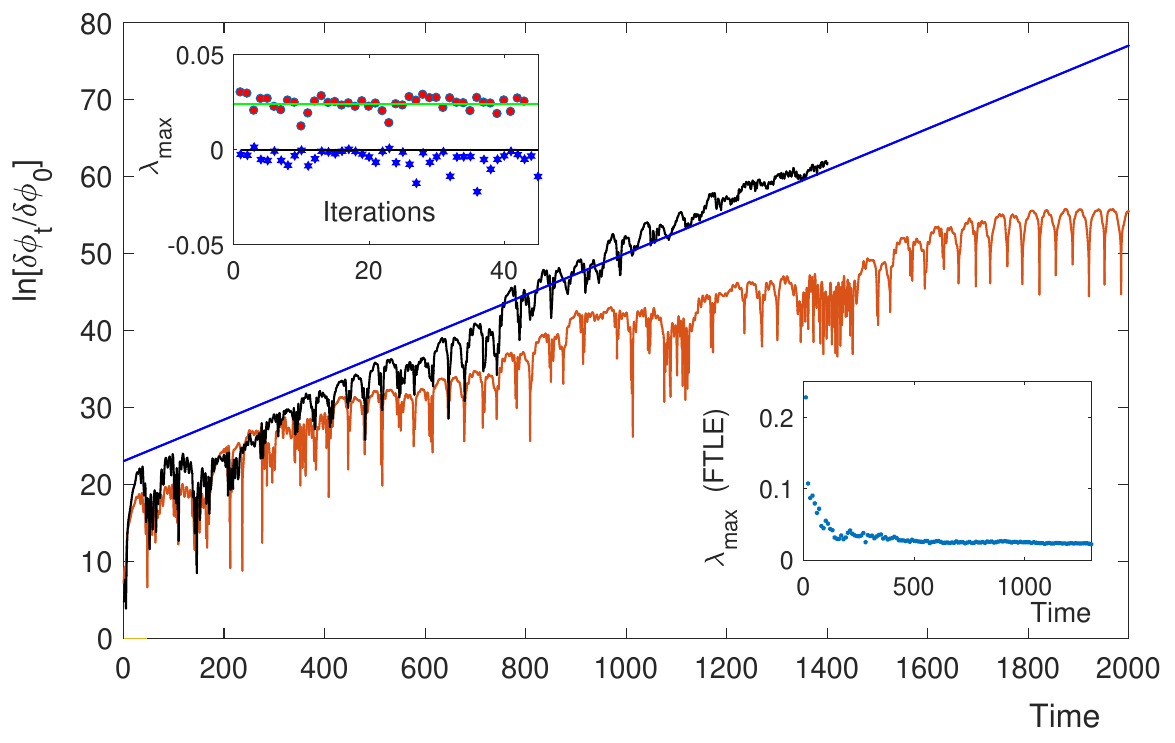}}
\caption{Main plot: Maximal Lyapunov exponent calculated by a simple method of monitoring evolution of infinitesimally close trajectories' deviation. Black line represents averaged trajectory differences for very close 200 trajectories Then fitting the inclination of the evolution of the function $ln\left(\delta\phi_t/\delta\phi_0\right)$ (blue straight line) gives the value of maximal Lyapunov exponent as $\lambda=0.026$. The upper inset shows the calculation of the maximal Lyapunov exponent by dividing the same chaotic trajectory by $N=45$ parts, filled red circles stand for the largest Lyapunov exponent of each part, while the horizontal straight line indicates the averaged value. Blue stars in the same graph, for comparison, present the same largest Lyapunov exponents, but now for a quasi-periodic regime. Lower inset depicts the dependence of FTLE on time working on the same chaotic trajectory. The initial conditions and dimensionless parameters are following: $\phi(0)=0.3$ $\ell(0)=0.05$, $\theta(0)=\pi/10$, All initial values of time derivatives $\dot\phi(0)$, $\dot\theta(0)$ and $\dot\ell(0)$ are zero and  $m_1/m_2=1.1$, $R/L=1/25$.} \label{fig6}
\end{figure}

\section{Quasi-Periodic and Chaotic Regimes}

As for quasi-periodic motion, first of all, it is realized that if we have reduced degrees of freedom of the system fixing initial $\theta=0$ and $\dot{\theta}=0$, those variables remain zero forever. Then one gets an oscillatory regime with an infinite lifetime shown in the upper graph of Fig. 2. In this case the system has a finite number of distinct oscillation periods, but it should be mentioned that this low dimensional case is full of chaotic trajectories, moreover, in this case, quasi-periodicity coexists with the chaos. this is a well-known and already fully investigated problem called "body swinging on a pulley" see ref.\cite{ragac} Paragraph [7.4.3]. While in the lower graph of Fig. 2 we present the case when $\theta(0)\neq 0$ for which the process seems chaotic and we are going to investigate this regime via different methods. We have measured the energy of the quasi-periodic case with initial $\theta$ and $\dot{\theta}$ equal to zero and then monitored the chaotic dynamics with the same energy for all the values with $\theta$ or $\dot{\theta}$ not equal to zero (see Fig.5).

On the other hand, choosing a nonzero initial value for $\theta$ we couldn't build some bifurcation diagram of transition to chaos: Besides the problems arising in conventional oscillatory systems, here we have one more related to “quickly ending” feature of the considered case. Indeed, even for an infinitesimal value of $\theta(0)$, system motion could be ended very quickly. Therefore, to distinguish between quickly ending and long-lasting regimes, we have plotted the graph which helps us to detect the point of transition from a quickly ending scenario towards the long lasting-chaotic regime. On the diagram presented in Fig. 3 the dependence of ending time (the time which is needed for $l_1$,$l_2$ or $l_3$ to reach zero value) on the initial angle of first pendulum $\phi(0)$ is depicted, where we fix other initial values as follows: $\ell(0)=0.05$, $\theta(0)=\pi/10$,  (all initial velocities are zero) and change initial value of $\phi$. In all our simulations we choose parameters $m_2/m_1=1.05$ and $r=1/25$. We also constructed the graph which shows how many trajectories survive at some specific time, based on this graph we have calculated one of the crucial quantity "escape rate" which measures how quickly trajectories escape from any neighborhood of the non-attracting chaotic set. The "escape time" sensitively depends on $\phi(0)$, as one can see in Fig. 3. At the ultimate left and right ends of the graph the oscillation lifetime is linearly dependent on $\phi(0)$ but between those two regions points are chaotically distributed on a graph and we show that this is the range where the chaos in the system takes place. As we have already mentioned this system has three stopping conditions when escape from the long-lasting dynamics takes place, we colored each escape condition on the graph to clearly show which one of these three conditions has more effect on the system's dynamics. As far as in this initial values range the process lasts for a long time, we can proceed with drawing Poincar\'e sections and calculating the largest Lyapunov exponents. 
The whole phase space is six-dimensional, thus for drawing Poincar\'e sections we collect the values of variables choosing a definite value for $\dot{\ell}=0$, then we construct the map in the plane of $\dot\phi$ versus $\phi$, shown in Fig. 4 for quasi-periodic and chaotic motions. As we have mentioned above, for quasi-periodic motion, first of all, we have fixed $\theta=0$ and change $\phi(0)$ keeping the initial velocities equal to zero (see different color curves corresponding to various initial angles $\phi(0)$ in the upper graph of Fig. 4). Increasing initial angle $\theta(0)$ and keeping $\phi(0)=0.3$ we still stay in the quasi-periodic regime (largest Lyapunov exponent is negative) and corresponding Poincar\'e sections are displayed in the middle graph of Fig. 4. This quasi-periodic regime survives until $\theta(0)=0.18$.  as we have already mentioned this system is high dimensional unlike the "body swinging on a pulley" problem \cite{ragac} that is the reason of intersections of quasi-periodic trajectory lines when $\theta>0$ as they are projections from a higher dimensional space onto the plane. For larger values of $\theta(0)$ and still keeping $\phi(0)=0.3$ the dynamics becomes chaotic and in lower graph of Fig. 4 we have chosen the initial value $\theta(0)=\pi/10$. In all graphs of Fig. 4, the depicted trajectories are taken from the lifetime graph depicted in Fig. 3 (for those trajectories oscillation lasts for approximately 100000-time units).

Next, we calculate the largest Lyapunov exponent for quantitative characterization of specific chaotic trajectories, namely for those two trajectories which are initially infinitesimally close to each other and both of them do last sufficient time $t_0$ to collect the information about their deviation. We calculate largest Lyapunov exponent $\lambda_{max}$ by using two different methods \cite{c18,c19,c20}. The expression for $\lambda_{max}$ reads as follows:
\begin{equation}
\label{a12}
\lambda=\lim_{t\to t_0}~\lim_{\delta\phi_0\to 0} \frac{1}{t}\ln\frac{\delta\phi_t}{\delta\phi_0}
\end{equation}
where $\delta\phi_t$ is deviation of two trajectories after time $t$ and $\delta\phi_0$ is initial perturbation at $t=0$.
For calculation of maximal Lyapunov exponent, we monitor the evolution of difference in time of angle $\phi$ for two close trajectories. In the main plot of Fig. 5, one can see the evolution of $ln\left(\delta\phi_t/\delta\phi_0\right)$ in time and the tangent of the inclination angle of the adjusting line is the largest Lyapunov exponent. For that particular trajectory the value is $\lambda_{max}=0.026$. At the same time, to be more confident, we have calculated the largest Lyapunov exponent by another method which provides dividing of the same chaotic trajectory by $N$ parts, then the calculation of the largest Lyapunov exponent of each part and finally averaging it (see ref \cite{c19,c20}). In the upper inset of Fig. 5, one can see the averaging procedure of different $\lambda$ -s (horizontal straight line displays the averaged value). After comparing it with the previous result we came up with a good agreement between these two calculations as $\lambda_{max}=0.026$ calculated by the first method and $\lambda_{max}=0.024$ calculated by the second method.

In order to include not only long-lasting regimes in our quantitative analysis, we also apply here Finite Time Lyapunov Exponent calculations as well. For this let us define three component deviation vector after time $t$ as $\delta{\bf X}_t\equiv(\delta\ell_t,\delta\phi_t,\delta\theta_t)$ with the main emphasis that its components are the functions of infinitesimal initial perturbations $\delta\ell_0$,  $\delta\phi_0$ and $\delta\theta_0$, i.e. one can explicitly write $\delta{\bf X}_t\equiv\delta{\bf X}(t,\delta\ell_0,\delta\phi_0,\delta\theta_0)$. Then one has simple matrix connection $\delta{\bf X}_t={\hat{\cal L}}\cdot\delta{\bf X}_0$ between initial perturbations and final trajectory deviation, where

\begin{equation}
{\hat{\cal L}}=
  {\begin{bmatrix}
\frac{\delta\ell(t,\delta\ell_0,0,0)}{\delta\ell_0}~~~~~~~ 
\frac{\delta\ell(t,0,\delta\phi_0,0)}{\delta\phi_0}~~~~~~~ \frac{\delta\ell(t,0,0,\delta\theta_0)}{\delta\theta_0} \\
\\
\frac{\delta\phi(t,\delta\ell_0,0,0)}{\delta\ell_0}~~~~~~~ 
\frac{\delta\phi(t,0,\delta\phi_0,0)}{\delta\phi_0}~~~~~~~ \frac{\delta\phi(t,0,0,\delta\theta_0)}{\delta\theta_0} \\
\\
\frac{\delta\theta(t,\delta\ell_0,0,0)}{\delta\ell_0}~~~~~~~ 
\frac{\delta\theta(t,0,\delta\phi_0,0)}{\delta\phi_0}~~~~~~~ \frac{\delta\theta(t,0,0,\delta\theta_0)}{\delta\theta_0} \\
  \end{bmatrix}}. \nonumber
\end{equation}

Then according to \cite{haller} one should calculate maximum eigenvalue $\Lambda_{max}$ of the matrix ${\hat{\cal L}}^T{\hat{\cal L}}$ and Largest Lyapunov exponent could be identified as
\begin{equation}
    \lambda_{max}=\frac{1}{t}\ln\left(\sqrt{\Lambda_{max}}\right). \label{ftle}
\end{equation}

In the lower inset of Fig. 5, we display the time dependence of this value for the same long-lasting chaotic trajectory which shows maximal Lyapunov exponent $\lambda_{max}=0.025$ in long time scale run and see that already at $t=50$ the FTLE gives realistic values for $\lambda_{max}$. Thus the method could be applied for finite time processes and we show in lower graphs of Fig. 1 the maximal Lyapunov exponent map for the wide range of initial values of $\ell$, $\phi$, and $\theta$. It should be emphasized that for all initial values one has positive values for the maximal Lyapunov exponent.

\section{Conclusions}

Quasi-periodic and chaotic features are fully investigated by tools such as Poincar\'e sections, Largest Lyapunov characteristic exponents, and Finite-Time Lyapunov Exponent method for the fallen pendula model. They reveal Quasi-periodic motion when $\theta=0$ and for small inclination angles of $\phi$. The system becomes chaotic in all cases except marginal ones with small initial deviation angles. This has been proven by the calculation of Finite-Time Lyapunov exponents (FTLE) and we believe that using the FTLE method for other oscillatory systems with a finite lifetime will be an effective tool for the investigation of transition points from regular to chaotic motion.

\bmhead{Acknowledgments}
We would like to thank T. Gachechiladze, G. Bakhtadze, G. Khomeriki and M. Osmanov-Baisera for interesting discussions and valuable comments.

\bmhead{Author contribution statement}: L. Osmanov contributed in setting general idea and implementation its numerical simulations as well as in writing manuscript draft. R. Khomeriki participated in choosing the methods for problem treatment and presentation of the simulation results as well as writing the paper draft.

\bmhead{Data Availability Statement}: The datasets generated during and/or analysed during the current study are available from the corresponding author on reasonable request.

\end{document}